\documentclass[]{spie}  %>>> use for US letter paper
\usepackage{amsmath,amsfonts,amssymb}
\usepackage{graphicx}

\usepackage[colorlinks=true, allcolors=blue]{hyperref}
\usepackage{amssymb}% http://ctan.org/pkg/amssymb
\usepackage{pifont}% http://ctan.org/pkg/pifont
\newcommand{\cmark}{\ding{51}}%
\newcommand{\xmark}{\ding{55}}%

\title{Invertible Residual Network with Regularization \\ for Effective Volumetric Segmentation}

\author[]{Kashu Yamazaki}
\author[]{Vidhiwar Singh Rathour}
\author[]{T. Hoang Ngan Le}
\affil[]{Department of Computer Science and Computer Engineering, \\ University of Arkansas, Fayetteville, Arkansas USA 72701.}
\authorinfo{Correspondence to T. Hoang Ngan Le\\ E-mail:thile@uark.edu, Telephone: (479) 575-6197}

\begin{document} 
\maketitle
\begin{abstract}
% 250 words for technical review
Deep Convolutional Neural Networks (CNNs) i.e. Residual Networks (ResNets) have been used successfully for many computer vision tasks, but are difficult to scale to 3D volumetric medical data. Memory is increasingly often the bottleneck when training 3D Convolutional Neural Networks (CNNs). Recently, invertible neural networks have been applied to significantly reduce activation memory footprint when training neural networks with backpropagation thanks to the invertible functions that allow retrieving input from its output without storing intermediate activations in memory to perform the backpropagation. 

Among many successful network architectures, 3D Unet \cite{C2016} has been established as a standard architecture for volumetric medical segmentation. Thus, we choose 3D Unet as a baseline for a non-invertible network and we then extend it with the invertible residual network. In this paper, we proposed two versions of invertible Residual Network, namely \textbf{Partially Invertible Residual Network (Partially-InvRes)} and \textbf{Fully Invertible Residual Network (Fully-InvRes)}. In Partially-InvRes, the invertible residual layer is defined by a technique called additive coupling \cite{NICE2015} whereas in Fully-InvRes, both invertible upsampling and downsampling operations are learned based on squeezing (known as pixel shuffle) \cite{squeeze}. Furthermore, to avoid the overfitting problem because of less training data, a variational auto-encoder (VAE) branch is added to reconstruct the input volumetric data itself. Our results indicate that by using partially/fully invertible networks as the central workhorse in volumetric segmentation, we not only reduce memory overhead but also achieve compatible segmentation performance compared against the non-invertible 3D Unet. We have demonstrated the proposed networks on various volumetric datasets such as iSeg 2019\cite{wang2019benchmark} and
BraTS 2020 \cite{brats_dataset}. 
\end{abstract}. 
% Deep Convolutional Neural Networks (CNNs) i.e. Residual Networks have been used successfully for many computer vision tasks, but are difficult to scale to 3D volumetric medical data. Memory is increasingly often the bottleneck when training 3D CNNs.  Among many successful network architectures, 3D Unet has been established as a standard architecture for volumetric medical segmentation. Thus, we choose Unet as a baseline for a non-invertible network and we then extend it with invertible residual network. In this paper, we proposed two versions of invertible Residual Network, namely Partially Invertible Residual Network and Fully Invertible Residual Network. To avoid the overtting problem because of less training data, a variational auto-encoder branch is added to reconstruct data itself. Our results indicate that we not only reduce memory overhead, but also achieve compatible segmentation performance compared against the non-invertible 3D Unet network.
%\newpage

\section{Introduction}
In recent years, deep learning-based methods, i.e. Convolutional Neural Networks (CNNs) have shown a great potential in medical imaging and archived state-of-the-art performance to solve challenging tasks such as detection \cite{breast_detection}, classification\cite{Disease_classification}, tracking \cite{cell_tracking}, and segmentation \cite{3Dunet, VNet, deepmedic, nonewnet, Brats_autoencoder}. Although deep neural networks have become a dominant method and archived high accuracy close to human performance for many computer vision tasks on 2D images, it is still challenging and limited when applying to medical tasks on volumetric data, such as volumetric segmentation, due to the limited amount of labelled data as well as the limited computational resources for training the model. While volumetric data is popular in biomedical imaging analysis, segment a volumetric is pixel-level annotation and very expensive. Based on the dimensions of convolutional kernel and input size, approaches for volumetric segmentation can be categorized into two: (i) 2D CNNs approaches and (ii) 3D CNNs approaches. The 2D CNNs can efficiently reduce the computational cost during training procedure but their performances are limited compared to the 3D CNNs approaches. Compare to 2D CNNs, the 3D CNNs have archived the state-of-the-art results in volumetric segmentation \cite{3Dunet, VNet, deepmedic, nonewnet, Brats_autoencoder} but their high computational cost becomes a bottleneck for the training of those models. Especially for large-scale data, most 3D CNN networks have prohibitive memory requirements.

In this paper, we propose two strategies, namely, partially invertible residual networks (Partially-InvRes) and fully invertible residual networks (Fully-InvRes to address the aforementioned problems in 3D CNNs for volumetric segmentation. Our goal is to improve the existing 3D CNNs networks (3D Unet is chosen in our case) in order to train a deeper and larger 3D network under the limited GPU memory requirement. 

\section{Proposed Methods}

In this work, we propose two versions of invertible residual networks, namely, partially invertible residual networks (Partially-InvRes) and fully invertible residual networks (Fully-InvRes). Our proposed networks make use of invertible residual layers proposed in \cite{ReversibleNet}. From the empirical results, it shows that invertible residual layers are very memory-efficient because intermediate activations do not have to be stored to perform backpropagation. During the backward pass, input activations that are required for gradient calculations can be computed from the output activations because the inverse function is accessible. This results in a constant spatial complexity ($\mathcal{O}$) in terms of layer depth. Table \ref{tab:revnet} shows the comparison between invertible network against other memory-efficient networks in the case a feed-forward network consisting of $L$ identical layers, the cost of forward propagation or backpropagation through a single layer is 1, and the memory cost of storing a single layer’s activations is 1. 

\begin{table*}[!h]
\centering
\caption{Comparison between Invertible Networks and other state-of-the-art memory-efficient networks}
\begin{tabular}{|l|l|l|}
\hline
Method            & \begin{tabular}[c]{@{}l@{}}Spatial \\ Complexity\\  (Activations)|\end{tabular} & \begin{tabular}[c]{@{}l@{}}Computational \\ Complexity\end{tabular} \\ \hline
Naive             & $\mathcal{O}(L)$      & $\mathcal{O}(L)$ \\ \hline
Checkpointing \cite{Checkpoint}    & $\mathcal{O}(\sqrt{L})$  & $\mathcal{O}(L)$ \\ \hline
Recursive  \cite{Recurrent}       &  $\mathcal{O}(log{L})$  & $\mathcal{O}(L \text{log}(L))$ \\ \hline
Invertible Networks \cite{ReversibleNet}& $\mathcal{O}(\sqrt{1})$  & $\mathcal{O}(L)$  \\ \hline
\end{tabular}
\label{tab:revnet}
\end{table*}

\subsection*{Partially Invertible Residual Networks (Partially-InvRes)}
\begin{figure*}[!h]
	\centering
	\includegraphics[width= 0.8\linewidth]{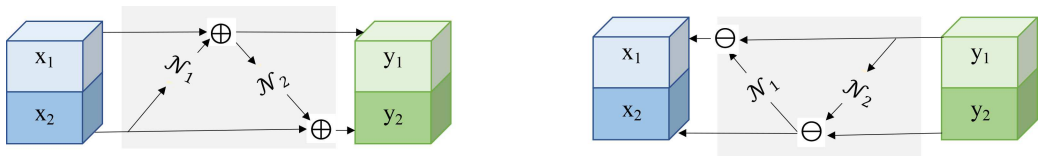}
	\caption{Illustration of forward (left), and the inverse (right) computations of an invertible block ${\mathcal{N}_1, \mathcal{N}_2}$ and its relationship with input $x_1, x_2$ together output $y_1, y_2$} 
	\label{fig:revnets}
\end{figure*}

In Partially-InvRes model, we obtain an invertible residual layer by the additive coupling \cite{NICE2015}, namely, partition the units in each layer into two groups. Let denotes $x_1$ and $x_2$ are two groups in each unit (channel). The invertible block contains two functions $\mathcal{N}_1, \mathcal{N}_2$ and the additive coupling rules between input $x_1, x_2$ and output $y_1, y_2$ is shown as in Eq.\ref{eq:invertible_block}. Fig.\ref{fig:revnets} gives an illustration of invertible block $\mathcal{N}_1, \mathcal{N}_2$ and its relationship with input $x_1, x_2$ together output $y_1, y_2$. 
\begin{equation}
\begin{split}
    y_1  = x_1 + \mathcal{N}_1(x_2) \quad  \quad \quad  &  x_1 = y_1 - \mathcal{N}_1(x_2) \\
    y_2 = x_2 + \mathcal{N}_2(y_1) \quad  \quad  & x_2 = y_2 - \mathcal{N}_2(y_1)
\end{split}
\label{eq:invertible_block}
\end{equation}

\begin{figure*}[!t]
	\centering
	\includegraphics[width= 0.85\linewidth]{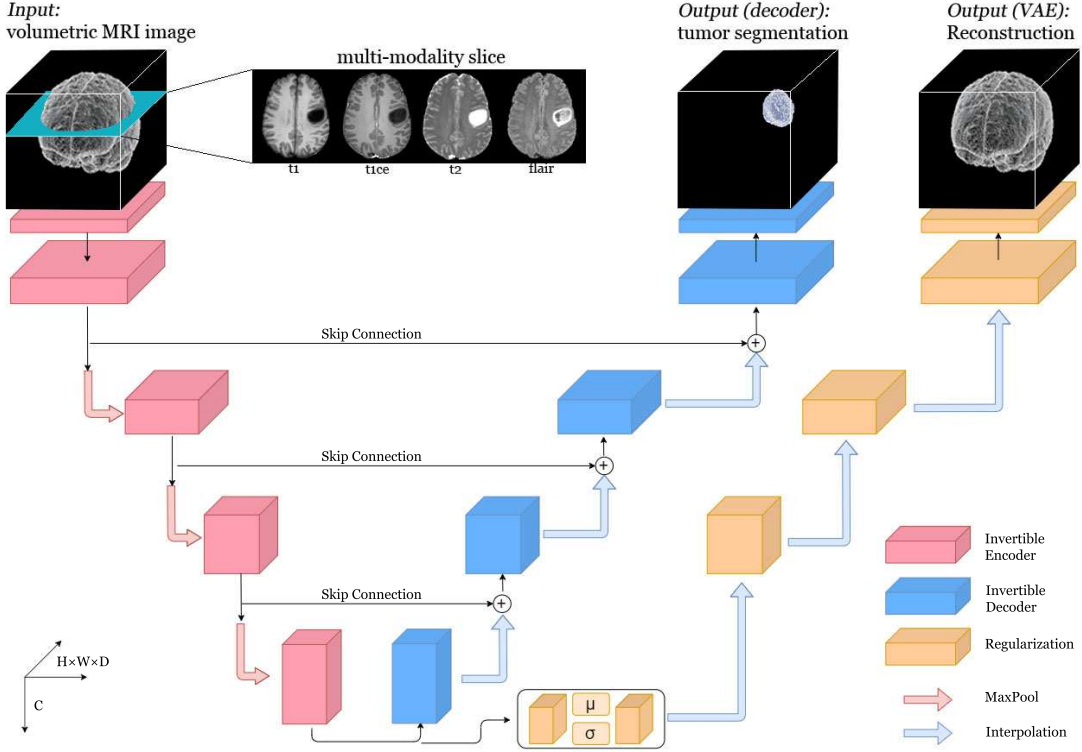}
	\caption{Illustration of our proposed Partially-InvRes with VAE regularization architecture.} 
	\label{fig:Partially-InvRes}
\end{figure*}

In order to avoid the overfitting problem as well as increase the network generalization, we added a variational autoencoder (VAE) regularization branch to reconstruct the input itself. The VAE regularization is employed by the Unet\_VAE network \cite{Myronenko2018}. The entire network architecture of Partially-InvRes is given in Fig. \ref{fig:Partially-InvRes}. In this figure, the Unet architecture is chosen as a baseline and we extend it with partially invertible layer and VAE regularization. 

As shown in Eq.\ref{eq:invertible_block} the number of input units is equal to the number of output units, thus invertible block can not be applied at downsampling and upsampling layers. In Partially-InvRes architecture, the invertible blocks are applied inside the encoders and decoders feature maps before downs- and upsampling layers in the encoder and decoder paths of Unet. The max pooling is applied to the downsampling layer and interpolation is applied to the upsampling layer. Thus, Partially-InvRes is invertible at feature map before down- and upsampling layers but it is not invertible at down- and upsampling layers.

\subsection*{Fully Invertible Residual Networks (Fully-InvRes)}

\begin{figure*}[!t]
	\centering
	\includegraphics[width= 0.85\linewidth]{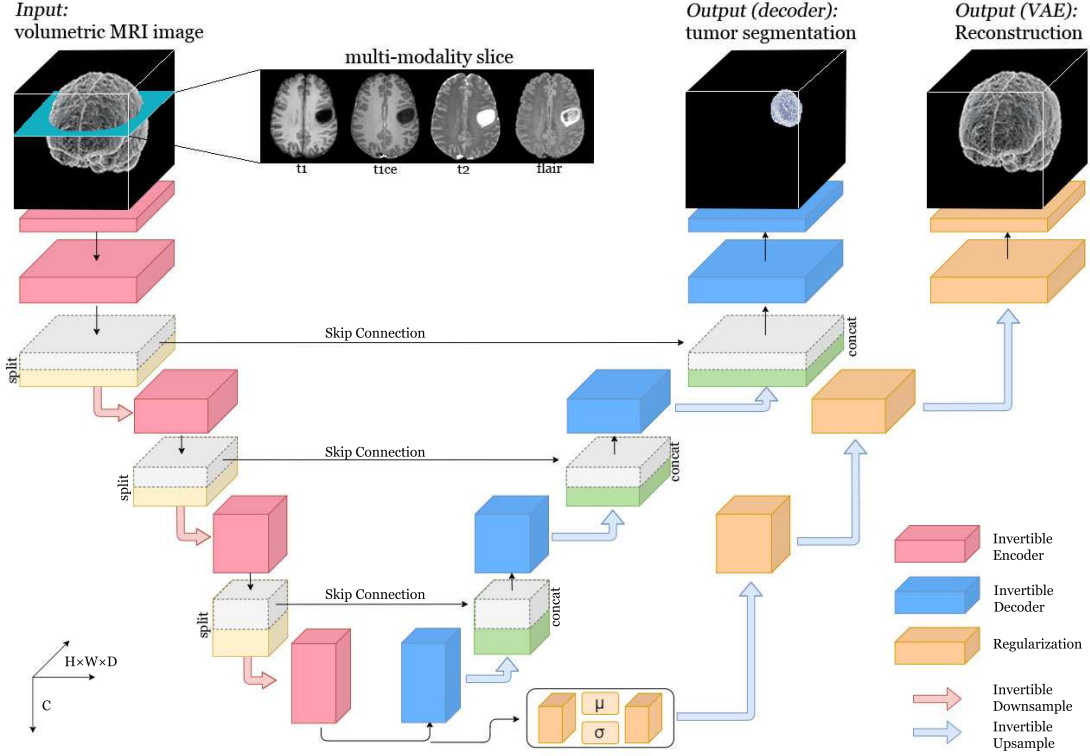}
	\caption{Illustration of our proposed Fully-InvRes with VAE regularization architecture.} 
	\label{fig:Fully-InvRes}
\end{figure*}

Different from Partially-InvRes which is not invertible at downsampling and upsampling layers, Fully-InvRes utilizes pixel shuffle or squeezing \cite{squeeze} so that both downsampling and upsampling layers are invertible. In Fully-InvRes, an input size $W\times H\times C$ is arranged into $\frac{W}{2}\times \frac{H}{2}\times 4C$ at downsampling layer. That means, it increases the number of channels ($C$)at the same time as it decreases the spatial resolution ($W, H$) of each channel. Opposite to the downsampling, in the upsampling, the number of channels needs to be decreased and the spatial resolution of each channel is increased. Fig. \ref{fig:Fully-InvRes} illustrated of Fully-InvRes architecture with three paths corresponding to Unet encoder path, Unet decoder path, and regularization path. In this network architecture, the learnable invertible downsampling layer is at the Unet encoder path and the learnable invertible upsampling is at the decoder path. At the Unet encoder path, each feature map is split along the channel direction into two parts. One part is invertible downsampling by pixel shuffle while the other part is transferring to the Unet decoder path through skip connections. At the Unet decoder path, each feature map is a combination of two parts. One part is from an invertible upsampling feature map which is conducted by pixel shuffle and the other part is from the skip connection. The AVE regularization \cite{Myronenko2018} is utilized to avoid the overfitting problem. 

We train our networks with a weighted combination of Dice loss ($\mathcal{L}_{Dice}$), Cross-entropy loss ($\mathcal{L}_{CE}$), L2 loss ($\mathcal{L}_{L2}$), and Kullback-Leibler divergence loss ($\mathcal{L}_{KL}$) as follows:
\begin{equation}
     \mathcal{L}_{} =  \mathcal{L}_{CE}+ \mathcal{L}_{Dice} + 0.1 \mathcal{L}_{L2}+ 0.1 \mathcal{L}_{KL}
\end{equation}
In this equation, $\mathcal{L}_{Dice}$ and $\mathcal{L}_{CE}$ are losses at the Unet decoder path whereas  $\mathcal{L}_{L2}$ and $\mathcal{L}_{KL}$ are losses at the AVE regularization path. 

\begin{itemize}
    \item \textbf{Dice loss ($\mathcal{L}_{Dice}$):} measures the degree of overlapping between the groundtruth $T$ and segmenting result $S$, sized of $W \times H$. Dice loss \cite{Milletari_2016} is come from Dice score which was used to evaluate the segmentation performance and defined as:
    \begin{equation}
        \mathcal{L}_{Dice} = 1 - 2\frac{\sum_i^{W \times H}{T_i S_i}}{\sum_i^{W \times H}{T_i+S_i}} = 1 - 2\frac{T \cap S}{T \cup S}
    \end{equation}
    \item \textbf{Cross Entropy loss ($\mathcal{L}_{CE}$):} is a widely used pixel-wise measure to evaluate the performance of classification or segmentation model. For binary segmentation, CE loss is expressed as Binary-CE (BCE) loss function as follows:
    \begin{equation}
        \mathcal{L}_{CE} = -\frac{1}{W \times H}\sum_{i=1}^{W \times H}{[T_iln(S_i) + (1 - T_i)ln(1-S_i)] }
    \end{equation}
    \item \textbf{Kullback-Leibler divergence divergence loss ($\mathcal{L}_{KL}$):} represents a loss based on Kullback-Leibler divergence between the estimated normal distribution $\mathcal{N}(\mu, \sigma^2)$ and a prior distribution $\mathcal{N}(0, 1)$ of $N$ image voxels:
    \begin{equation}
         \mathcal{L}_{KL} = \frac{1}{N}\sum(\mu^2+\sigma^2-\log \sigma^2 -1)
    \end{equation}
    \item \textbf{L2 loss ($\mathcal{L}_{L2}$): }is computed on regularization branch between output reconstruction volumetric $V_{recons}$ and the input volumetric $V_{input}$:
    \begin{equation}
        \mathcal{L}_{2} = ||V_{recons} - V_{input}||^2
    \end{equation}

\end{itemize}

\section{Results \& Comparison}

\subsection{Dataset}
Following volumetric datasets are used to conduct the experiments:

\textbf{iSeg:} The iSeg19 dataset \cite{wang2019benchmark} consists of 10 subjects with ground-truth labels for training and 13 subjects without ground-truth labels for testing. Each subject includes T1 and T2 images with size of $144 \times 192 \times 256$, and image resolution of $1 \times 1 \times 1$  $\text{mm}^3$. In iSeg, there are three classes:  white matter (WM), gray matter (GM), and cerebrospinal fluid (CSF). We use 9 subjects for training and 1 subject for testing.

\textbf{BraTS:} The BraTS 2020 database  \cite{Brats}contains 369 scans with labels. For each scan, there are 4 available modalities, i.e., T1, T1C, T2, and Flair. Each image is registered to a common space, sampled to an isotropic $1 \times 1 \times 1$ $\text{mm}^3$ resolution by the organizers and has a dimension of $240 \times 240 \times 155$. In BraTS 2020, there are three tumor classes: whole tumor (WT), tumor core (TC), and enhanced tumor (ET). We use 80\% data for training and 20\% data for testing. 

\subsection{Metrics}
The metrics used to benchmark our performance are as follows:

\textbf{Dice Score (DSC):} Dice score between the groundtruth $T$ and the segmenting result $S$ is computed as: 

\begin{equation}
DSC(S,T) = 2 \frac{T \cap S}{T \cup S}
\end{equation}

\textbf{Hausdorff distance (Hauf):} Hausdorff distance is an evaluation metric that calculates the distance between segmentation boundaries, i.e., the surface distance. Hausdorff distance is the maximum value of overall shortest least-squares distance $d$ calculated for all points $p$ on the surface $\delta P_1$ of a given volume $P_1$ to points $t$ on the surface $\delta T_1$ of the other given volume $T_1$, and vice versa.
\begin{equation}
    %\small
{\rm Hauf}(P,T) = \max\{\sup_{p \in \partial P_1} \, \inf_{t \in \partial T_1} d(p,t),\, \sup_{t \in \partial T_1} \, \inf_{p \in \partial P_1} d(t,p)\}
\end{equation}
%Hausdorff measure is highly sensitive to small outlying subregions(outliers) because it returns maximum over 'all' surface distances. 

\subsection{Experiment Setting} 
Our 3D architecture is built upon 3D-Unet \cite{3Dunet} and the input is defined as $N \times C \times H \times W \times D$, where $N$ is the batch size, $C$ is the number of input modalities, and $H, W, D$ are height, width and depth of volume patch on sagittal, coronal, and axial planes. We implemented our network using PyTorch 1.3.0 and our model is trained until convergence by using the ADAM optimizer. We employed the Adam optimizer, with a learning rate of 2e-4. Our 3D Unet makes use of instance normalization \cite{Instance_normalization} and Leaky ReLU.  The experiments are conducted using an Intel CPU, and RTX GPU. 

\subsection{Results}
Both Partially-InvRes and Fully-InvRes are implemented in PyTorch using the library MemCNN \cite{memCNN} for memory-efficient backpropagation. The experimental results have been conducted on two volumetric datasets including iSeg-2019\cite{iSeg_dataset} and BraTS 2020\cite{Brats}. Because iSeg-2019 contains a small number of samples, we randomly partition each volumetric into 1024 samples sized $64\times 64 \times 64$ whereas we use a patch size of $128\times 128 \times 128$ in BraTS 2020 dataset. We choose 3D Unet \cite{3Dunet} as a non-invertible baseline. The performance on iSeg-2019 and BraTS 2020 are given in Table \ref{tab:iseg}, \ref{tab:brats} respectively, where the best performance is highlighted in bold. From the empirical results, we can see that Partially-InvRes and Fully-InvRes achieve compatible segmentation performance in terms of Dice score and Hausdorff distance and they both are better than the baseline. However, in terms of memory that is required during training procedure, Fully-InvRes is better than Partially-InvRes and they both are better than the baseline. 

The table \ref{tab:memory_compare}shows the comparison between our proposed methods against other state-of-the-art networks in the terms of memory efficiency, overfitting problem, and segmentation performance.

\begin{table*}[!h]
\centering
\caption{Results on iSeg2019 validation set with the input size is set as $64\times 64\times 64$}
\begin{tabular}{|c|c|c|c|c|c|c|c|}
\hline
Method        & Memory & \multicolumn{3}{c|}{DSC} &  \multicolumn{3}{c|}{Hauf} \\ \hline
          &        & CSF     & GM     & WM       & CSF     & GM     & WM     \\ \hline
Baseline      &    4.18GB    &   92.50     &  90.84      &   89.26 &   9.45  &    6.96    &   7.47     \\ \hline
Partially-InvRes    &  3.60GB     &    95.07      &  \textbf{91.66 }     &   \textbf{90.92}    & 9.62 &     \textbf{6.54}   &    \textbf{6.87}      \\ \hline
Fully-InvRes     &    \textbf{2.12GB}   &   \textbf{95.09}      &   91.32     &   90.31     &    \textbf{9.48}    &   6.87     &      6.91          \\ \hline
\end{tabular}
\label{tab:iseg}
\end{table*}

% [*] Dice: (WT) 0.8597, (TC) 0.7469, (ET) 0.6970
% [*] Hausdorff Distance: (WT) 10.5673, (TC) 11.2846, (ET) 8.7111
% [*] Sensitivity: (WT) 0.8314, (TC) 0.7899, (ET) nan
% [*] Specificity: (WT) 0.9976, (TC) 0.9970, (ET) 0.9985 

% [*] Dice: (WT) 0.8820, (TC) 0.7774, (ET) 0.7201
% [*] Hausdorff Distance: (WT) 10.4670, (TC) 8.3307, (ET) 6.6887
% [*] Sensitivity: (WT) 0.8874, (TC) 0.7810, (ET) nan
% [*] Specificity: (WT) 0.9961, (TC) 0.9980, (ET) 0.9984 

\begin{table*}[!h]
\centering
\caption{Results on BraTS2020 validation set with the input size is set as $128\times 128\times 128$}
\begin{tabular}{|c|c|c|c|c|c|c|c|}
\hline
Method        & Memory & \multicolumn{3}{c|}{DSC} &  \multicolumn{3}{c|}{Hauf} \\ \hline
          &       & WT    & ET     & TC     & WT     & ET     & TC     \\ \hline
Baseline      &    6.8GB    &   84.68     &  \textit{70.16}     &  74.13   &   11.45    &  9.13   &   11.37    \\ \hline
%3D Unet \cite{3DUnet_baseline} & 8.75GB  &  82.86 & 66.92 & 72.98 & 17.37 & 44.96 & 28.24 \\ 
Partially-InvRes    &  5.7GB     &  \textit{85.97}    &  69.70  &  \textit{74.69}    &   \textit{10.56  }   &  \textit{8.71 } &   \textit{11.28}    \\ \hline
Fully-InvRes     &  \textbf{3.1GB }&  \textbf{88.20} &  \textbf{72.01 }  &   \textbf{77.74}     &  \textbf{10.46} &  \textbf{6.68} & \textbf{8.33}  \\ \hline 

\end{tabular}
\label{tab:brats}
\end{table*}

\begin{table}[!h]
\centering
\caption{Comparison between our proposed networks and state-of-the-art networks on volumetric segmentation}
\begin{tabular}{|l|l|l|l|}
\hline
Models & Memory Efficient& Avoid Overfiting & High Accuracy \\ \hline
3D Unet \cite{3Dunet} & \xmark  & \xmark & \cmark  \\ \hline
No New-Net \cite{nonewnet}       & \xmark  & \xmark        & \cmark      \\ \hline
Autoencoder Unet \cite{Brats_autoencoder} & \xmark     & \cmark        & \cmark    \\ \hline
Our Partially-InvRes    & \cmark  & \cmark        & \cmark  \\ \hline
Our Fully-InvRes    & \cmark  & \cmark        & \cmark  \\ \hline
\end{tabular}
\label{tab:memory_compare}
\end{table}

\section{Conclusions}
In this work, we introduced two different approaches to address the problem of memory footprint when training a high-dimensional setting as in medical volumetric image segmentation. In the first approach, Partially-InvRes, we obtain an invertible residual layer by the additive coupling \cite{NICE2015} while down-and upsampling layers are not invertible and performed by max pooling and interpolation, respectively. In the second approach, Fully-InvRes, both down-and upsampling layers are invertible and performed by pixel shuffle or squeezing \cite{squeeze}. We demonstrate our proposed approaches for memory-efficient training on the two medical datasets for volumetric segmentation. The experimental results have shown that our proposed methods Partially-InvRes and Fully-InvRes provide compatible segmentation performance and they both are better than the baseline while the memory usage in the proposed methods is much efficient than the baseline during training.

\newpage
\bibliography{main} % bibliography data in report.bib
\bibliographystyle{spiebib} % makes bibtex use spiebib.bst

\end{document}